\documentstyle[12pt,aaspp4,graphics]{article}

\begin{document}
\title{Time Dilation in the Peak-to-Peak Time Scale of GRBs}
\author{Ming Deng\altaffilmark{1}, and Bradley E. Schaefer\altaffilmark{2}
\altaffiltext{1}{ming.deng@yale.edu}
\altaffiltext{2}{schaefer@grb2.physics.yale.edu}
}
\affil{Department of Physics, Yale University, New Haven, CT 06520-8120}

\baselineskip 24pt
\clearpage

\begin{abstract}

We present strong evidence of time stretching in the peak-to-peak time
scales in the light curves of BATSE Gamma Ray Bursts (GRBs). Extensive 
tests are performed on artificially dilated bursts to verify that the 
procedure for extracting the
peak-to-peak time scales correctly recovers the stretching of bursts.
The resulting robust algorithm is then applied to the 4B GRB database. We
derive a stretching factor of $1.92 \pm 0.13$  between the brightest burst
group ($P > 7.7 photon \cdot cm^{-2} \cdot s^{-1}$) and the dimmest burst
group ($P = 1.0 \sim 1.4 photon \cdot cm^{-2} \cdot s^{-1}$) with several
independent peak-to-peak time scale definitions and they agree within
uncertainties. Such an agreement strongly supports the interpretation of
the observed time stretching as
time dilation caused by the cosmological expansion, rather than physical
selection effects. We fit the result to cosmological models with $\Lambda
= 0$, $\Omega_{0}$ from 0.2 to 1.0, and contrained the standard candle 
luminosity to be  $L_{0}= 7.0 \pm 2.0 \pm 2.7 \times 10^{56} photons \cdot
s^{-1}$. Our luminosity value is fully consistent with the value
from the combined PVO and BATSE LogN-LogP curve with the BATSE bright
bursts at low redshifts of $z_{bright} = 0.11 \pm 0.02 \pm 0.025$. 
This luminosity fit is definitely inconsistent with the the larger 
distance scale implied from associating burst density with star formation
rates. 

\end{abstract}


\clearpage                               
\section{Introduction}

Observations taken by the BATSE instrument aboard the Compton Gamma Ray
Observertory have identified $\sim 1800$ GRBs and shown that their angular 
distribution is highly isotropic, and their distribution in space is 
inhomogeneous (Meegan et al. 1992, Briggs et al. 1996). Such a
distribution naturally arises if GRBs are at cosmological distances.
The logN-LogP distribution of GRBs is well studied and shown
(e.g. Fenimore et al. 1993) to be consistent with cosmological models. The
recent measurement of the
redshift of GRB970508 (Metzger et al. 1997) between $0.835 \sim 2.1$, and
the possible identification of the host galaxy of GRB971214 (Kulkarni et
al. 1998) at $z \sim 3.42$ provide further evidence supporting the
cosmological scenario. 

Piran (1992) and Paczy\'{n}ski (1992) suggested that the light curves of
the 
GRBs should be stretched due to the cosmological time dilation. This
effect applies to all time scale in the GRB lightcurves. In the past, 
different groups (Norris et al. 1994; Mitrofanov et al. 1996; Bonnel et
al. 1996; Rutledge et al. 1996 ) have investigated the correlation of the
duration of bursts ($T_{50}$ , $T_{90}$, width of the main peak) and the
burst brightness (peak flux). The results have been contradictory, with
different groups getting different results using identical method and
similar data sets.
   
Furthermore, questions have been raised whether the time stretching 
found is due to intrinsic correlation (Brainerd 1994, 1997) between pulse
width and burst brightness for bursts drawn from a volume limited sample.
It is argued that the observed correlation could arise either from the
beaming of relativistic jets (Brainerd 1994), or the conservation of total
energy in monoenergetic sources (Wijers \& Paczy\'{n}ski 1994). Thus the
controversial claims of temporal stretching need neither imply
cosmological distances nor be of any utility in understanding burst
demographics.

In this Letter, we investigate the correlations of the time intervals
between peaks with brightness indicators peak flux P. The
peak-to-peak time scale is independent of earlier pulse width
measurements. Attempts (Pozanenko et al. 1997, Norris et al. 1996) have
been made to search for peaks separated by a valley with intensity
difference of at least $4 \sigma$. Such a definition is biased towards
identifying more peaks in the bright bursts than in the dimmer bursts
since the latter have less photon counts above background. We used
artificially stretched bursts to identify a peak-to-peak time scale which
correctly recovers the input stretching relations.    

\section{Peak Finder Algorithm}

We have used the updated BATSE 64ms ascii database which contains 
1252 bursts ranging from trigger 105 to trigger 5624. This database
provides 64ms time bins from the concatenation of three standard BATSE
datatypes, DISCLA, PREB and DISCSC. All three data types
are derived from the on-board data stream of BATSE's eight Large Area
Detectors (LADs), and
all three data types have four energy channels, with approximate channel
boundaries: $25-55 keV$, $55-110 keV$, $110-320 keV$, and $> 320 keV$
(see Fishman et al. 1989). The peak flux values are derived on the $256ms$
time scale as in the 4th BATSE catalog.

The data were binned to 256ms to achieve better S/N ratio for the dim
bursts. Bursts with $T_{90}$ shorter than 2s are excluded (Norris et
al. 1994, Mitrofanov et al. 1996) from
our analysis. The noise biases are rendered uniform (Norris et al. 1994) 
by diminishing the background subtracted signal to a canonical peak
intensity of $2400 counts \cdot s^{-1}$, and adding a canonical flat
background of $8000 counts \cdot s^{-1}$. The elimination of bursts
with peak flux below the $1.0 photon \cdot cm^{-2} \cdot s^{-1}$ threshold
is introduced to avoid trigger threshold effects and ensure the
statistical significance of the peaks identified.

The detailed procedure to find all the peaks in a burst is as follows:
(1) We fit the background using a quadratic function to the
pre- and postburst regions. The background is subtracted from the data.
(2) The maximum peak counting rate $C_{max}$ at 256ms is identified.  (3)
Identify all local maximums $C_{p}$ that are separated by local minimums
$C_{min}$ which satisfy $C_{p} - C_{min} > N_{v} \cdot C_{max}$. 
(4) Each such local maximum also has to be greater than a threshold
level, $C_{p} > T \cdot C_{max}$. $N_{v}$ is a parameter to
ensure the peaks are distinct enough statistically. $T$ is a
threshold level for accepting a local maximum as a significant peak. 

We used four definitions of peak-to-peak time scales. The first method
uses all time intervals between successive peaks, $\tau_{i}$. Thus a burst
with N peaks will provide N-1 peak-to-peak time scales, all of which have
equal weight. Our second method is to logarithmically average all the time
intervals between successive peaks $\tau_{p-p}$ for each individual burst,
resulting in one time scale per burst. Our
third method is to consider only the time interval $\tau_{HS}$ between the
highest and the second highest peak. The fourth method is to consider the
time interval $\tau_{FL}$ between the first and the last peaks identified. 

\section{Simulation Tests}

The intervals between peaks for long bursts ($T_{90} > 2s$) varies from
less than 1 second to tens of seconds, thus any dilation effect can be
detected only in a statistical sense. Furthermore, the identification
of peaks are complicated by small S/N ratios in weak bursts which might
introduce noise peaks. The highly significant peaks do not suffer from 
these problems, while having more peaks
identified will improve the statistics of our analysis. Therefore, it is
not known {\em a priori} which parameters $N_{v}$ and $T$ are the best
choice to measure the time intervals between peaks of GRBs, nor do we
have {\em a priori} knowledge of which peak-to-peak time scale definition
yields reliable results. A faithful procedure should be able to
distinguish the time stretching effect from any systematic effects.

To find a faithful procedure, we simulated the time stretching
effects of the bursts and tested whether our procedure of using the peak
finder algorithm in extracting the time scale correctly recovers the time
stretching despite the systematic effects. Our procedure is to search
the parameter space for $N_{v}$ and $T$ to find regions where the peak
finder algorithm recovers the input dilation with high confidence. The
details of the simulation procedure go as follows:
                                             
We assume a standard cosmology with $\Lambda = 0, \Omega_{0} = 2q_{0}$.
Hence there is the following relationship between peak flux and redshift.

\begin{equation}
P(z)=\frac{5.2\times 10^{-58} \cdot q_{0}^{4} \cdot
(1+z)^{-\alpha+2}\cdot
L}
{[zq_{0}+(q_{0}-1)(\sqrt{2q_{0}z+1}-1)]^{2}}
\end{equation}

Here $q_{0}$ is the deceleration parameter, $\alpha$ is the
power law spectral index of GRBs, $L$ is the burst
luminsity in $photons \cdot s^{-1}$. In deriving this relationship, we
assumed a Hubble constant of $75  km \cdot s^{-1} \cdot Mpc^{-1}$. The
power law spectral index varies from 1.0 to 3.0 for a majority of the 
bursts (Schaefer et al. 1994). We assumed $q_{0} = 0.5$, $\alpha = 2.0$,
and a typical (Horack et al. 1996, Hakkila et al. 1996) GRB power law
luminosity function $\phi(L) = A_{0}L^{-\beta}$ ($L_{min} < L < L_{max}$, 
$K = L_{max}/L_{min}=100.0$) with $\beta = 2.0, L_{min} = 2.3 \times
10^{57} photons \cdot s^{-1}$ in the simulation process of stretched
bursts. 

With this cosmology, we created a database of simulated BATSE light
curves. Each burst was simulated to have peak flux value P randomly
sampled from the $LogN - LogP$ relation of the 4B catalog, and its
luminosity $L$ randomly sampled from the luminosity function $\phi(L)$
assumed. Both of them are sampled using the rejection method (Press et
al. 1992).
Its redshift $z$ is then determined by Eq. 1. We then randomly select a
burst from a library
of 100 bright BATSE bursts, and determined its redshift $z_{bright}$
from the above relation. Burst photon counts from the 4 different channels
from the Large Area Detectors are redistributed by a smooth interpolation
function to redshift the photon energy. The energy shift factor is 
$\epsilon = (1+z_{bright})/(1+z)$.

	For the burst at redshift z to be simulated, the relative time
dilation factor is $S = \epsilon ^{-1} = (1+z)/(1+z_{bright})$.
The light curve is then stretched by a factor of S. The photon 
count of $ith$ bin $N_{i}$ in the stretched light curve is dependent upon
the photon count of two bins $O_{p}$ and $O_{p+1}$ in the original light
curve, where $p$ is the integer portion of $i/S$. The fraction of
$O_{p+1}$ that will contribute is $f = i - p \cdot S$ ($f = 1.0$ if $i - p
\cdot S > 1.0$). Taking account of the new background, the new count
$N_{i}$ should be:

\begin{equation}
N_{i} = O_{p+1} \cdot f  + O_{p} \cdot (1-f) + p(d),
d = O_{p+1} \cdot f^{2} + O_{p} \cdot (1-f)^{2}
\end{equation}

where p(d) is a Poisson variable whose standard deviation is $d^{1/2}$.
The resulting light curve is then dimmed by a factor of 
$P(z)/P(z_{bright})$ as prescribed by Norris et al. 1994.   

	This process is repeated 1250 times to generate a simulated database 
of gamma ray burst light curves, with the same $Log N-Log P$ distribution
as the real 4B GRB database. Each database simulation is repeated 10 times
with different seeds for random number generation. By looking at the
scatter between these 10 simulations we estimate the uncertainty
associated with the simulation procedure.

	The peak finder algorithm is applied to each simulated database.
We divide the bursts into 6 brightness bins with equal numbers of
bursts, and logarithmically averaged the individual peak-to-peak
times within each bin. The average peak-to-peak time scales derived are
normalized to that of the brightest bin. The resulting relation is then
compared to the input dilation relationship of Eq. 1 by a $\chi^{2}$
test. The $\chi^{2}$ values were calculated for a wide range of $N_{v}$
and $T$. 

In Fig. 1, the $\chi^{2}$ dependency on $N_{v}$ and $T$ is shown by a
contour plot for the simulation with the stretching relation of Eq. 1. We
also performed simulations for bursts with no stretching, the
$\chi^{2}$ dependency on  $N_{v}$ and $T$ is shown in Fig. 2. 
The contour areas that reside within $\chi^{2} = 4$ in each figure
provide $N_{v}$ and $T$ values that define a peak-to-peak time scale
which faithfully recovers the input relation. Obviously, such areas are
much larger when we have no stretching (Fig. 2). The overlapping
areas ($N_{v} \sim 0.37, T < 0.22$) provide a definition that
reproduce the dilation relation with high confidence for both the
stretched and unstretched simulated database. We repeated the simulation 
with a variety of cosmological parameters and luminosity values and found
that such a conclusion consistently holds. In Table 1, the time
stretching relation recovered from the algorithm is compared to the input
relation. These simulation procedures ensure that we have a peak-finding
algorithm which faithfully returns the stretching factors regardless of
the size of stretching.

	The primary parameter that ensures the peaks found are
nonstatistical is $N_{v}$. It is defined to be the peak-to-valley
amplitude (measured in units of the highest peak height above background)
separating each candidate peak from its neighbors. When $N_{v}$ is small, 
the algorithm picks up many statistical fluctuations as peaks, which
distort the stretching relation. When $N_{v}$ is large, very few peaks are
identified, and the statistics are poor. Nevertheless, these few peaks
will result in large uncertainties if they are used as a measurement of
$\tau_{i}$ such that $\tau_{i}$ vs. $P$ relation can be fitted to any 
model with a reasonable $\chi^{2}$. In figures 1 and 2, this is evidenced 
by a region of $\chi^{2} < 6$ across the top for large $N_{v}$ values. 
While faithful, these regions carry little information as to distinguish 
between models. Therefore the algorithm only works for intermediate
$N_{v}$ values.

	The parameter $T$ in the algorithm is defined to be a threshold 
level that all peaks identified have to exceed. When $N_{v} > T$, the $T$
constraint is primarily important to reject some large statistical
fluctuations above
and below the background. Although rare, such fluctuations usually yield 
a peak in the background data far away from the burst activity, creating
large $\tau_{i}$ values that are false and distort the apparent $\tau_{i}$
vs. $P$ relation.

\section{Results and Conclusions}

	We applied the algorithm to the real BATSE database, and the
result is shown in Fig. 3. A stretching factor $S$ is calculated by
fitting the resulting $\tau vs. P$ relationship to Equation 1. The same
procedure for simulation test is repeated for the other definitions of
peak-to-peak time scales and the resulting algorithm is applied to the
real BATSE bursts. The results are summarized in Table 2. The time
stretching factor $S$ of $1.92 \pm 0.13$ between the brightest ($P > 7.7
photon \cdot cm^{-2} \cdot s^{-1}$) and dimmest group ($P = 1.0 \sim 1.4
photon \cdot cm^{-2} \cdot s^{-1}$) is found. Within the uncertainties,
the different definitions of the peak-to-peak time scales give consistent
results. 

	Recently, two spectral class of bursts have been identified
(Pendleton et al. 1997): those bursts with high energy emission (HE), and
those with no high energy (NHE) emission. The latter group is
predominantly fainter than HE bursts and is found to have a homogeneity to
much lower flux values. We applied the peak finder algorithm (see Table 3)
to this subgroup with 232 bursts in the 3B catalog. Since the NHE bursts
are inherently much fainter, there are fewer bursts with peak flux P above
the $1.0 photon \cdot cm^{-2} \cdot s^{-1}$ threshold (66 out of 232) and
hence the uncertainties are much larger. The time stretching factors of
the NHE bursts appears to be consistent with those of the HE bursts
(reduced $\chi ^{2}_{5}=1.01$) and those of a no stretching model (reduced
$\chi ^{2}_{5}=1.11$). Therefore, it is safe to conclude that the limited 
statistics available is not enough to determine whether any stretching
exist in NHE bursts. 
 	
	It should be noted that the time dilation relation is independent
of the GRB density evolution. We fit the result to the Eq. 1
and constrain the luminosity value $L_{0}$ in a standard candle scenario.
This provides an estimate of the redshift $z_{bright}$ of brightest BATSE
bursts independent of previous estimates from the $Log N-Log P$ relation.
We fit the data with varying values of deceleration parameters
$q_{0}$ from  $0.1$ to  $0.5$, and power index of GRB spectra $\alpha$
from $1.0$ to $2.0$. We find the best fitting luminosity $L_{0} =
7.0 \pm 2.0 \pm 2.7 \times 10^{56} photons \cdot s^{-1}$ (reduced
$\chi ^{2}_{4}=1.23$), and the brightest bursts (peak flux $P = 30 photons 
\cdot cm ^{-2} s^{-1}$) at $z_{bright} = 0.11 \pm 0.02 \pm 0.025$.
The first error denotes the uncertainty associated with the varying values
of $q_{0}$ and $\alpha$, the second error is derived from the $\chi^{2}$
fits. This result favors the simple cosmological scenario of bright
bursts at small redshifts rather than evolutionary scenario of bursts at
much larger redshifts based on some theoretical arguments (Totani 1997,
Paczy\'{n}ski 1997) that burst density traces star formation rates.

	It is well known that GRBs might have a broad luminosity function.
We performed our simulation adopting a range dominated luminosity model
(Horack 1996, Hakkila 1996) and demonstrated that for typical luminosity
function $\phi(L) \propto L^{-\beta}$ ($L_{min} < L < L_{max}$, $K =
L_{max}/L_{min}=100.0$) with
power-law index $\beta$ near 2, the broadening of the luminosity
distribution will not smear the time dilation effect out. Our simulation
shows that our technique can correctly recover the time stretching on a
bright burst sample for this case. However, the time dilation data alone
can not be used to constrain all the parameters $L_{min}, L_{max}, \beta$
in the power law luminosity function as it involves too many free
parameters in the cosmological model.  

	In summary, we have measured several peak-to-peak time scales of
GRBs. Simulation tests are performed to verify that the time scales identified 
correctly reflect the physical time scale intrinsic
to the bursts. The $\tau_{HS}$ and $\tau_{FL}$ time scales are similar to 
the durations, while the $\tau_{i}$ and $\tau_{p-p}$ peak-to-peak time
scales measurements are independent of previous duration and pulse width
measurements. Our result unambiguously shows the existence of time
stretching factor of $s = 1.92 \pm 0.13$ between the bright and dimmest 
GRB time profiles, and the agreement found between these multiple
independent time scale measurements indicates that the temporal profiles
of GRBs are universally stretched. Such an agreement implies that the time
stretching relation comes from cosmological expansion rather than physical
selection effects affecting particular property of bursts. 

	We thank G. N. Pendleton for providing the list of NHE bursts.
This work has made use of the data obtained though the Compton Gamma Ray
Observatory Science Support Center Online Service, provided by the Goddard 
Space Flight Center.

\clearpage

\begin {table}
\begin {center}
\begin {tabular}{|c|c|c|}
\hline
$P(photons \cdot cm^{-2} \cdot s^{-1})$ & $s_{No dilation}$ &
$s_{recovered}$ \\
\hline
$1.0 \sim 1.4$ & 1.00 & $ 1.04 \pm 0.10$  \\ 
$1.4 \sim 1.8$ & 1.00 & $ 1.05 \pm 0.09$  \\ 
$1.8 \sim 2.5$ & 1.00 & $ 1.13 \pm 0.09$  \\ 
$2.5 \sim 3.6$ & 1.00 & $ 0.98 \pm 0.08$  \\
$3.6 \sim 7.7$ & 1.00 & $ 1.10 \pm 0.09$  \\
$> 7.7$        & 1.00 & $ 1.00 \pm 0.09$  \\
\hline
$P(photons \cdot cm^{-2} \cdot s^{-1})$ & $s_{Dilation}$ &
$s_{recovered}$ \\
\hline
$1.0 \sim 1.4$ & 1.69 & $ 1.82 \pm 0.12$ \\
$1.4 \sim 1.8$ & 1.56 & $ 1.67 \pm 0.09$ \\
$1.8 \sim 2.5$ & 1.44 & $ 1.58 \pm 0.11$ \\
$2.5 \sim 3.6$ & 1.30 & $ 1.28 \pm 0.12$ \\
$3.6 \sim 7.7$ & 1.17 & $ 1.12 \pm 0.11$ \\
$> 7.7       $ & 1.00 & $ 1.00 \pm 0.08$ \\
\hline
\end{tabular}
\end{center}
\caption{The input stretching factor and the measured factor using the
definition of peak-to-peak time scales $<\tau_{p-p}>$ with $N_{v} =
0.375$ and $T = 0.100$ for both simulated stretched bursts and unstretched
bursts.}

\end{table}

\begin {table}
\begin {center} 
\begin {tabular}{|c|c|c|}
\hline
Peak-to-peak Timescale & Stretching Factor\\
Definition & $s_{observed}$ \\
\hline
$\tau_{i}$   & $1.92 \pm 0.13$\\
\hline
$<\tau_{p-p}>$ & $1.96 \pm 0.17$\\
\hline
$\tau_{HS}$  & $1.92 \pm 0.26$\\
\hline
$\tau_{FL}$  &  $1.92 \pm
0.20$\\
\hline
\end {tabular}
\end{center}
\caption{The observed time stretching factor between the brightest burst
group ($P > 7.7 photon \cdot cm^{-2} \cdot s^{-1}$) and the dimmest
burst
group ($P = 1.0 \sim 1.4 photon \cdot cm^{-2} \cdot s^{-1}$) using
different definition of time scales.}
\end{table}

\begin {table}
\begin {center}
\begin {tabular}{|c|c|c|c|}
\hline
$P(photons \cdot cm^{-2} \cdot s^{-1})$ & $s_{NHE}$ & $s_{HE}$ &
$s_{All}$ \\
\hline
$1.0 \sim 1.4$ & $ 1.68  \pm 0.49$ & $1.83 \pm 0.23$ & $1.92 \pm 0.24$ \\
$1.4 \sim 1.8$ & $ 2.97  \pm 1.28$ & $1.56 \pm 0.22$ & $1.65 \pm 0.20$ \\
$1.8 \sim 2.5$ & $ 0.90  \pm 0.28$ & $1.40 \pm 0.18$ & $1.27 \pm 0.16$ \\
$2.5 \sim 3.6$ & $ 1.54  \pm 0.60$ & $1.24 \pm 0.14$ & $1.42 \pm 0.16$ \\
$3.6 \sim 7.7$ & $ 1.40  \pm 0.70$ & $1.03 \pm 0.12$ & $1.08 \pm 0.13$ \\
$> 7.7       $ & $ 1.00  \pm 0.29$ & $1.00 \pm 0.10$ & $1.00 \pm 0.09$ \\
\hline
\end{tabular}
\end{center}
\caption{The stretching factor relative to the brightest burst group for
NHE bursts, HE bursts and all the bursts.  Notice that the uncertainties
associated with the NHE bursts are substantially larger due to the
relative scarcity of such events in the 4B catalog.}
\end{table}

\begin{figure}
\begin{center}
\resizebox{7cm}{7cm}{\includegraphics{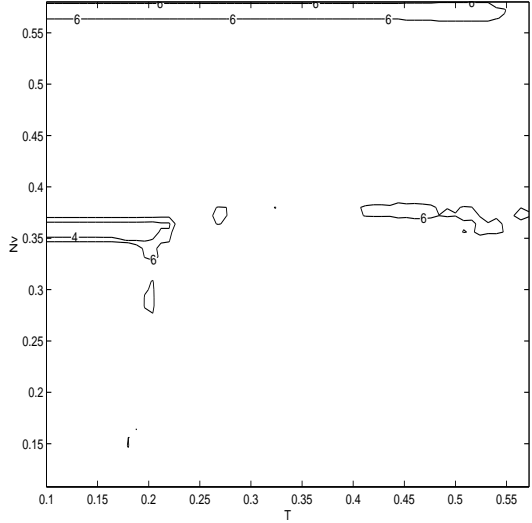}}
\caption{The $\chi^{2}$ contour of the peak-to-peak algorithm for
cosmological model with $L_{0} = 2.3 \times 10^{57} photons \cdot s^{-1}$ 
. There are two level of $\chi^{2}$s, 4 and 6.
The x and y axis are $T$ and $N_{v}$ values respectively. The $\chi^{2}$ 
values are calculated by fitting $\tau vs. P$ relation recovered by the
algorithm to the input stretching relation of 6 burst brightness groups.
There are 4 degrees of freedom for the fit, hence the region within
$\chi^{2} = 4$ provides a faithful algorithm which correctly recovers the 
input stretching relation.
}
\end{center}
\end{figure}

\begin{figure}
\begin{center}
\resizebox{7cm}{7cm}{\includegraphics{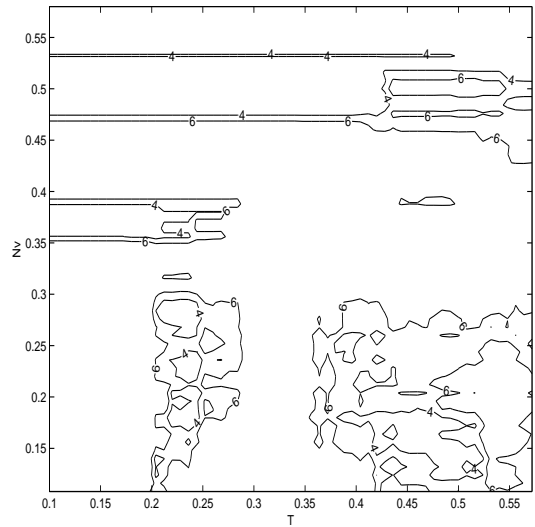}}
\caption{The $\chi^{2}$ contour of the peak-to-peak algorithm for
no dilation model. There are two level of $\chi^{2}$s, 4 and 6.
The x and y axis are $T$ and $N_{v}$ values respectively. The $\chi^{2}$ 
values are calculated by fitting $\tau vs. P$ relation recovered by the
algorithm to the input stretching relation of 6 burst brightness groups.
There are 4 degrees of freedom for the fit, hence the region within
$\chi^{2} = 4$ provides a faithful algorithm. Notice that the faithful
areas are much larger in this case.
}
\end{center}
\end{figure}

\begin{figure}
\begin{center}
\resizebox{15cm}{12cm}{\includegraphics{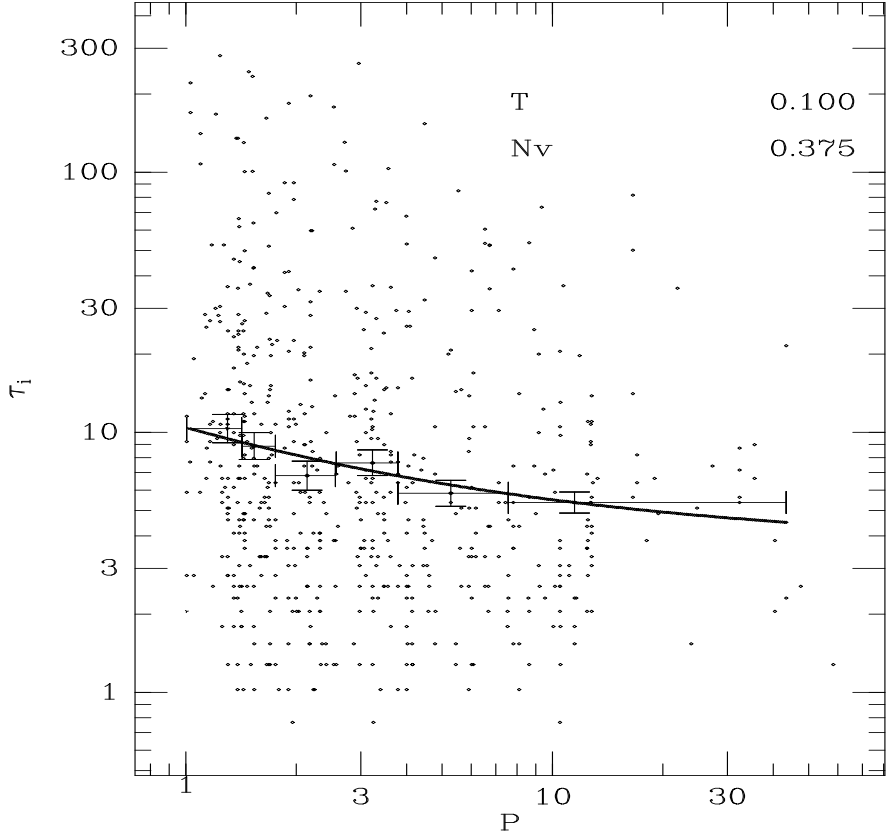}}
\caption{The average peak-to-peak intervals of BATSE GRB database. Each
dot in the
graph represents the peak-to-peak interval identified $\tau_{i}$ in a
mutipeak burst. P is the peak flux in $photons \cdot cm^{-2} \cdot
s^{-1}$. $\tau_{i}$ is the peak-to-peak interval in units of seconds. The
solid curve is the best fit to a cosmological model with $q_{0} = 0.2$,
$\Lambda = 0$, $\alpha = 1.5$, and burst standard candle luminosity 
$L_{0} = 7.0 \pm 2.7 \times 10^{56} photons \cdot s^{-1}$. }
\end{center}
\end{figure}

\end{document}